\documentclass{article}

\usepackage{epsfig}
\usepackage{named}
\usepackage{amsmath,theorem}
\usepackage{xspace}
\usepackage{amssymb}
\usepackage{times,euscript}

  \theoremstyle{break}
  \theoremheaderfont{\bf\boldmath}
  \newtheorem{satz}{Satz}[section]
  \newtheorem{theorem}[satz]{Theorem}
  \newtheorem{definition}[satz]{Definition}
  \theoremstyle{plain}
  \newtheorem{lemma}[satz]{Lemma}

\newcommand{\qed}{{\unskip\nobreak\hfil\penalty50
   \hskip2em\hbox{}\nobreak\hfil
   \qedd
   \parfillskip=0pt \finalhyphendemerits=0
    \medskip\goodbreak\noindent}}
\newcommand{\qedd}{\vrule height4pt width 4pt depth0pt}

\newfont{\mymathtt}{cmtt10 scaled 1095}

\newcommand{\alc}{
  \ensuremath{\mathcal{ALC}}\xspace}

\newcommand{\alcRtrans}{
        \ensuremath{\alc_{R^{+}}}\xspace}

\newcommand{\s}{
        \ensuremath{\mathcal{S}}\xspace}

\newcommand{\shiq}{
  \ensuremath{\mathcal{SHIQ}}\xspace}

\newcommand{\A}{\ensuremath{\mathcal{A}}\xspace}

\newcommand{\Fact}{FaCT\xspace}
\newcommand{\PDL}
        {PDL\xspace}
\newcommand{\CPDL}
        {\emph{converse}-PDL\xspace}

\newfont{\bigmathxx}{cmsy10 scaled 1440}
\newfont{\smallmathxx}{cmsy10 scaled 720}

\newcommand{\bigsqcap}{\mathop{\mathop{\mbox{\bigmathxx\symbol{117}}}}\limits}

\newcommand{\I}{
        \ensuremath{\mathcal{I}}\xspace}

\newcommand{\ifunc}{
        \ensuremath{^\mathcal{I}}\xspace}
\newcommand{\deltai}{
        \ensuremath{\Delta\ifunc}\xspace}

\newcommand{\Term}{
        \ensuremath{\mathcal{T}}\xspace}

\newcommand{\Concepts}{\ensuremath{\mathbf{C}}\xspace}
\newcommand{\Roles}{\ensuremath{\mathbf{R}}\xspace}
\newcommand{\Inds}{\ensuremath{\mathbf{I}}\xspace}

\newcommand{\Rplus}{\ensuremath{\mathbf{R}_+}\xspace}

\newcommand{\Lab}{\ensuremath{\EuScript{L}}\xspace}

\newcommand{\Edges}{\ensuremath{\EuScript{E}}\xspace}
\newcommand{\Inv}{\mathop{\mathsf{Inv}}}
\newcommand{\Tr}{\mathop{\mathsf{Trans}}}

\newcommand{\K}
        {\ensuremath{\mathbf{K}}\xspace}
\newcommand{\KT}
        {\ensuremath{\mathbf{KT}}\xspace}
\newcommand{\Kfour}
        {\ensuremath{\mathbf{K4}}\xspace}
\newcommand{\Sfour}
        {\ensuremath{\mathbf{S4}}\xspace}
\newcommand{\Km}
        {\ensuremath{\K_{(\mathbf{m})}}\xspace}
\newcommand{\KTm}
        {\ensuremath{\KT_{(\mathbf{m})}}\xspace}
\newcommand{\Kfourm}
        {\ensuremath{\mathbf{K4}_{(\mathbf{m})}}\xspace}
\newcommand{\Sfourm}
        {\ensuremath{\mathbf{S4}_{(\mathbf{m})}}\xspace}

\newcommand{\con}[1]{
        \ensuremath{\textsf{#1}}}

\newcommand{\cname}[1]{\con{#1}}

\newcommand{\Tail}{\mathop{\mathsf{Tail}}}

\newcommand{\some}[2]{%
        \ensuremath{\exists #1 . #2}}
\newcommand{\all}[2]{%
        \ensuremath{\forall #1 . #2}}

\newcommand{\atleastq}[3]{%
        \ensuremath{\mbox{$\geqslant$}#1 #2 . #3}}
\newcommand{\atmostq}[3]{%
        \ensuremath{\mbox{$\leqslant$}#1 #2 . #3}}

\newcommand{\tuple}[2]{
        \ensuremath{\langle #1 , #2 \rangle}}

\newcommand{\Card}[1]{
        \ensuremath{\vert#1\vert}}

\newcommand{\sss}{{\mathrel{\kern.25em{\sqsubseteq}\kern-.5em \mbox{{\scriptsize
*}}\kern.25em}}}
\newcommand{\ssse}{{\mathrel{\kern.25em{\sqsubseteq}\kern-.5em \mbox{{\scriptsize
*}}\kern.25em}_1}}
\newcommand{\sssz}{{\mathrel{\kern.25em{\sqsubseteq}\kern-.5em \mbox{{\scriptsize
*}}\kern.25em}_2}}

\newcommand{\R}{\ensuremath{\mathcal{R}}\xspace}

\title{Reasoning with Individuals for the Description Logic \shiq \thanks{This
    paper will appear in the Proceedings of the 17th International Conference
    on Automated Deduction (CADE-17), Lecture Notes in Computer Science,
    Germany, 2000. Springer Verlag. }}

\author{{\bf Ian Horrocks}\\
  Department of Computer Science, University of Manchester\\
  \texttt{horrocks@cs.man.ac.uk} \and 
  {\bf Ulrike Sattler}\\
  LuFG Theoretical Computer Science, RWTH Aachen\\
  \texttt{sattler@informatik.rwth-aachen.de} \and 
  {\bf Stephan Tobies}\\
  LuFG Theoretical Computer Science, RWTH Aachen\\
  \texttt{tobies@informatik.rwth-aachen.de}}

\date{}

\newcommand{\path}[1]{[#1]}
  
\newcommand{\Indmap}{\ensuremath{\EuScript{I}}\xspace}
\newcommand{\clos}{\mathsf{clos}}
\newcommand{\ndoteq}{\not\doteq}
\newcommand{\nneg}{\ensuremath{\mathord{\sim}}}
\newcommand{\pair}[2]{\frac{#1}{#2}}
\newcommand{\Paths}{\ensuremath{\mathsf{Paths}}\xspace}

\newcommand{\tabl}[1]{(\textsf P#1)}
\newcommand{\btabl}[1]{\textbf{(\textsf P#1)}}
\newcommand{\for}{\ensuremath{\mathcal{F}}\xspace}

\begin{document}
\maketitle

\begin{abstract}
While there has been a great deal of work on the development
of reasoning algorithms for expressive description logics, in most
cases only Tbox reasoning is considered. In this paper we present an
algorithm for combined Tbox and Abox reasoning in the \shiq
description logic.  This algorithm is of particular interest as it can
be used to decide the problem of (database) conjunctive query
containment w.r.t.\ a schema. Moreover, the realisation of an
efficient implementation should be relatively straightforward as it
can be based on an existing highly optimised implementation of the
Tbox algorithm in the \Fact system.
\end{abstract}


\newcommand{\DLR}{\ensuremath{\mathcal{DLR}}\xspace}

\section{Motivation}

A description logic (DL) knowledge base (KB) is made up of two parts, a
terminological part (the terminology or Tbox) and an assertional part (the
Abox), each part consisting of a set of axioms. The Tbox asserts facts
about \emph{concepts} (sets of objects) and \emph{roles} (binary relations),
usually in the form of inclusion axioms, while the Abox asserts facts about
\emph{individuals} (single objects), usually in the form of instantiation
axioms. For example, a Tbox might contain an axiom asserting that
\cname{Man} is subsumed by \cname{Animal}, while an Abox might contain axioms
asserting that both \cname{Aristotle} and \cname{Plato} are instances of the
concept \cname{Man} and that the pair \tuple{\cname{Aristotle}}{\cname{Plato}}
is an instance of the role \cname{Pupil-of}.

For logics that include full negation, all common DL reasoning tasks are
reducible to deciding KB consistency, i.e., determining if a given KB admits a
non-empty interpretation~\cite{Buchheit93}.  There has been a great deal of work
on the development of reasoning algorithms for expressive
DLs~\cite{Baad90c,GiLe95,Horrocks98j,DeGiaMass98}, but in most cases these
consider only Tbox reasoning (i.e., the Abox is assumed to be empty).  With
expressive DLs, determining consistency of a Tbox can often be reduced to
determining the satisfiability of a single
concept~\cite{Baad90c,Schi91,BBNNS93}, and---as most DLs enjoy the tree model
property (i.e., if a concept has a model, then it has a tree model)---this
problem can be decided using a tableau-based decision procedure.

The relative lack of interest in Abox reasoning can also be explained by the
fact that many applications only require Tbox reasoning, e.g.,
ontological engineering~\cite{Horrocks96a,Mays96a} and schema
integration~\cite{CDLNR98}. Of particular interest in this regard is the DL
\shiq~\cite{Horrocks99j}, which is powerful enough to encode the logic
\DLR~\cite{CDLNR98}, and which can thus be used for reasoning about conceptual
data models, e.g., Entity-Relationship (ER) schemas~\cite{Calvanese98c}.
Moreover, if we think of the Tbox as a \emph{schema} and the Abox as
(possibly incomplete) \emph{data}, then it seems reasonable to assume that
realistic Tboxes will be of limited size, whereas realistic Aboxes could
be of almost unlimited size.  Given the high complexity of reasoning in most
DLs~\cite{Schi91,Calvanese96a}, this suggests that Abox reasoning could lead to
severe tractability problems in realistic applications.\footnote{Although
suitably optimised algorithms may make reasoning practicable for quite large
Aboxes~\cite{Haarslev99b}.}

However, \shiq Abox reasoning is of particular interest as it allows \DLR schema
reasoning to be extended to reasoning about conjunctive query containment
w.r.t.\ a schema~\cite{CaldGL-PODS-98}.  This is achieved by using Abox
individuals to represent variables and constants in the queries, and to enforce
co-references~\cite{HorrocksSattler+-LTCS-99-15}.  In this context, the size of
the Abox would be quite small (it is bounded by the number of variables
occurring in the queries), and should not lead to severe tractability problems.

Moreover, an alternative view of the Abox is that it provides a restricted form
of reasoning with \emph{nominals}, i.e., allowing individual names to appear in
concepts~\cite{Schaerf94,Blackburn98a,Areces99a}.  Unrestricted nominals are
very powerful, allowing arbitrary co-references to be enforced and thus leading
to the loss of the tree model property.  This makes it much harder to prove
decidability and to devise decision procedures (the decidability of \shiq with
unrestricted nominals is still an open problem). An Abox, on the other hand, can
be modelled by a \emph{forest}, a set of trees whose root nodes form an
arbitrarily connected graph, where number of trees is limited by the number of
individual names occurring in the Abox. Even the restricted form of
co-referencing provided by an Abox is quite powerful, and can extend the range
of applications for the DLs reasoning services.

In this paper we present a tableaux based algorithm for deciding the
satisfiability of unrestricted \shiq KBs (i.e., ones where the Abox may be
non-empty) that extends the existing consistency algorithm for
Tboxes~\cite{Horrocks99j} by making use of the forest model property.
This should make the realisation of an efficient implementation relatively
straightforward as it can be based on an existing highly optimised
implementation of the Tbox algorithm (e.g., in the \Fact
system~\cite{Horrocks99c}). A notable feature of the algorithm is that, instead
of making a unique name assumption w.r.t.\ all individuals (an assumption
commonly made in DLs~\cite{Baader91b}), increased flexibility is provided by
allowing the Abox to contain axioms explicitly asserting inequalities between
pairs of individual names (adding such an axiom for every pair of individual
names is obviously equivalent to making a unique name assumption).


\section{Preliminaries}

\renewcommand{\sss}{{\mathrel{\kern.25em{\sqsubseteq}\kern-.5em
\raisebox{0ex}{{\scriptsize *}}\kern.25em}}}
\newcommand{\mycolon}{\! :\!}

In this section, we introduce the DL \shiq. This includes the definition of
syntax, semantics, inference problems (concept subsumption and satisfiability,
Abox consistency, and all of these problems with respect to
terminologies\footnote{We use \emph{terminologies} instead of
Tboxes to underline the fact that we allow for general concept inclusions axioms
and do not disallow cycles.}), and their relationships.

\shiq is based on an extension of the well known DL \alc~\cite{SSSm91} to
include transitively closed primitive roles~\cite{Sat95ashort}; we call this
logic \s due to its relationship with the proposition (multi) modal logic
\Sfourm~\cite{Schi91}.\footnote{The logic \s has previously been called
\alcRtrans, but this becomes too cumbersome when adding letters to represent
additional features.} This basic DL is then extended with inverse roles
($\mathcal{I}$), role hierarchies ($\mathcal{H}$), and qualifying number
restrictions ($\mathcal{Q}$).

\begin{definition}\label{syntax}
  Let $\Concepts$ be a set of \emph{concept names} and \Roles a set of
  \emph{role names} with a subset $\Rplus\subseteq \Roles$ of \emph{transitive
  role names}.
  The set of {\em roles\/} is $\Roles \cup \{R^-\mid R\in\Roles \}$. To avoid
  considering roles such as $R^{--}$, we define a function $\Inv$ on roles such
  that $\Inv(R) = R^-$ if $R$ is a role name, and $\Inv(R) = S$ if $R=S^-$.  We
  also define a function $\Tr$ which returns $\mathrm{true}$ iff $R$ is a
  transitive role. More precisely, $\Tr(R) = \mathrm{true}$ iff $R\in \Rplus$ or
  $\Inv(R)\in\Rplus$.
  
  A \emph{role inclusion axiom} is an expression of the form $R\sqsubseteq S$,
  where $R$ and $S$ are roles, each of which can be inverse. A \emph{role
  hierarchy} is a set of role inclusion axioms.
  For a role hierarchy \R, we define the relation $\sss$ to be the
  transitive-reflexive closure of $\sqsubseteq$ over $\R\cup\{
  \Inv(R)\sqsubseteq \Inv(S)\mid R\sqsubseteq S \in \R\}$. A role $R$ is called
  a \emph{sub-role} (resp. super-role) of a role $S$ if $R \sss S$ (resp. $S
  \sss R$).
  A role is \emph{simple} if it is neither transitive nor has any transitive
  sub-roles.

  The set of \shiq-\emph{concepts} is the smallest set such that
  \begin{itemize}
  \item every concept name is a concept, and,
  \item if $C$, $D$ are concepts, $R$ is a role, $S$ is a simple role, and $n$
    is a nonnegative integer, then $C\sqcap D$, $C\sqcup D$, $\neg C$, $\forall
    R.C$, $\exists R.C$, $\atleastq n S C$, and $\atmostq n S C$ are also
    concepts.
  \end{itemize}
    A \emph{general concept inclusion axiom} (GCI) is an expression of the form
  $C\sqsubseteq D$ for two \shiq-concepts $C$ and $D$. A \emph{terminology} is a
  set of GCIs.
  
  Let $\Inds =\{a,b,c\ldots \}$ be a set of \emph{individual names}. An
  \emph{assertion} is of the form $a\mycolon C$, $(a,b)\mycolon R$, or $a
  \ndoteq b$ for $a,b\in \Inds$, a (possibly inverse) role $R$, and a
  \shiq-concept $C$. An \emph{Abox} is a finite set of assertions.
\end{definition}  
  
Next, we define  semantics of \shiq and the corresponding inference problems. 

\begin{definition}\label{semantics}
  An {\em interpretation\/} $\I = (\Delta^\I,\cdot^\I)$ consists of a set
  $\Delta^\I$, called the {\em domain\/} of $\I$, and a \emph{valuation}
  $\cdot^\I$ which maps every concept to a subset of $\Delta^\I$ and every role
  to a subset of $\Delta^\I\times\Delta^\I$ such that, for all concepts $C$,
  $D$, roles $R$, $S$, and non-negative integers $n$, the following equations
  are satisfied, where $\sharp M$ denotes the cardinality of a set $M$ and $
  (R\ifunc)^+$ the transitive closure of $R\ifunc$:

  \noindent\begin{tabular}{rclr}
    $R\ifunc $&$=$&$ (R\ifunc)^+$ & for each role $R\in \Rplus$ \\
    $(R^-)\ifunc$ &$=$& $\{\tuple{x}{y} \mid \tuple{y}{x} \in R\ifunc\} $ &
    (inverse roles)\\
    $(C \sqcap D)\ifunc$ &$=$& $C\ifunc \cap D\ifunc$ & (conjunction)\\
   $(C \sqcup D)\ifunc$ &$=$& $C\ifunc \cup D\ifunc$ & (disjunction)\\
   $(\neg C )\ifunc$ &$=$& $ \Delta\ifunc \setminus C\ifunc$ & (negation)\\
   $(\some{R}{C})\ifunc$&$=$ & $\{x \mid \exists y.\tuple{x}{y} \in R\ifunc \mbox{ and
   }y \in C\ifunc\}$  & (exists restriction)\\
   $(\all{R}{C})\ifunc$&$=$ & $\{x \mid \forall  y.\tuple{x}{y} \in R\ifunc \mbox{
   implies }y \in C\ifunc\}$  & (value restriction)\\
   $(\atleastq{n}{R}{C})\ifunc$&$=$& $\{x \mid \sharp \{y.\tuple{x}{y} \in
   R\ifunc \mbox{ and } y \in C\ifunc\} \geqslant n\} $& ($\geqslant$-number
   restriction)\\
   $(\atmostq{n}{R}{C})\ifunc$&$=$& $\{x \mid \sharp \{y.\tuple{x}{y} \in
   R\ifunc \mbox{ and } y \in C\ifunc\} \leqslant n\} $& \hspace{0.5cm}($\leqslant$-number restriction)\\
   
  \end{tabular}
  
  An interpretation $\I$ \emph{satisfies} a role hierarchy $\R$ iff $R^\I \subseteq
  S^\I$ for each $R \sqsubseteq S$ in $\R$. Such an interpretation is called a
  \emph{model} of \R (written  $\I \models \R$).
  
  An interpretation $\I$ \emph{satisfies} a terminology $\Term$ iff $C^\I\subseteq
  D^\I$ for each GCI $C\sqsubseteq D$ in \Term.  Such an interpretation
  is called a \emph{model} of \Term (written $\I \models \Term$).
  
  A concept $C$ is called {\em satisfiable\/} with respect to a role hierarchy
  $\R$ and a terminology $\Term$ iff there is a model $\I$ of $\R$ and $\Term$
  with $C^\I \neq \emptyset$.
  A concept $D$ {\em subsumes\/} a concept $C$ w.r.t. $\R$ and $\Term$ iff $C^\I
  \subseteq D^\I$ holds for each model $\I$ of $\R$ and $\Term$.  For an
  interpretation $\I$, an element $x \in \Delta^\I$ is called an {\em instance}
  of a concept $C$ iff $x\in C^\I$.

  For Aboxes, an interpretation maps, additionally, each individual $a\in \Inds$
  to some element $a^\I\in \Delta^\I$. An interpretation $\I$ satisfies an
  assertion
  \begin{center}
    $\begin{array}[t]{rcl}
      a\mycolon C &\mbox{ iff } & a^\I\in C^\I, \\
      (a,b)\mycolon R & \mbox{ iff } & \tuple{a^\I}{b^\I}\in R^\I, \mbox{ and}\\
      a\ndoteq b & \mbox{ iff } & a^\I \neq b^\I
    \end{array}$
  \end{center}
  An Abox $\A$ is \emph{consistent} w.r.t. $\R$ and $\Term$ iff there is a model
  $\I$ of $\R$ and $\Term$ that satisfies each assertion in $\A$.
\end{definition}
For DLs that are closed under negation, subsumption and (un)satisfiability can
be mutually reduced: $C \sqsubseteq D$ iff $C\sqcap \neg D$ is unsatisfiable,
and $C$ is unsatisfiable iff $C\sqsubseteq A\sqcap \neg A$ for some concept name
$A$.  Moreover, a concept $C$ is satisfiable iff the Abox $\{a\:C\}$ is
consistent. It is straightforward to extend these reductions to role
hierarchies, but terminologies deserve special care:
In \cite{Baad90c,Schi91,BBNNS93}, the \emph{internalisation} of GCIs is
introduced, a technique that reduces reasoning w.r.t. a (possibly cyclic)
terminology to reasoning w.r.t. the empty terminology. For \shiq, this reduction
must be slightly modified. The following Lemma shows how general concept
inclusion axioms can be \emph{internalised} using a ``universal'' role $U$, that
is, a transitive super-role of all roles occurring in \Term and their respective
inverses.

\begin{lemma}\label{lemma:terminologies}
  Let $C,D$ be concepts, \A an Abox, \Term a terminology, and \R a role
  hierarchy. We define
  $$C_\Term:= \bigsqcap_{C_i\sqsubseteq D_i \in\Term}\neg C_i\sqcup D_i.$$
  \noindent  Let $U$ be a transitive role that does not occur in $\Term$, $C$, $D$, $\A$, or $\R$. We
  set
  $$\R_U := \R \cup \{ R\sqsubseteq U, \Inv(R)\sqsubseteq U \mid \text{$R$
  occurs in $\Term$, $C$, $D$, \A, or $\R$} \}.$$
  \begin{itemize}
  \item $C$ is satisfiable w.r.t. \Term and $\R$ iff
    $C\sqcap C_\Term \sqcap \all{U}{C_\Term}$
    is satisfiable w.r.t. $\R_U$.
  \item $D$ subsumes $C$ with respect to \Term and $\R$ iff
    $C\sqcap\neg D\sqcap C_\Term \sqcap \all{U}{C_\Term}$
    is unsatisfiable w.r.t. $\R_U$.
  \item \A is consistent with respect to \R and \Term iff 
    $\A\cup\{a\mycolon C_\Term \sqcap \all{U}{C_\Term} \mid a \mbox{ occurs in \A
    } \}$
    is consistent w.r.t. $\R_U$.
  \end{itemize}
\end{lemma}

The proof of Lemma~\ref{lemma:terminologies} is similar to the ones
that can be found in \cite{Schi91,Baad90c}. Most importantly, it must
be shown that, (a) if a \shiq-concept $C$ is satisfiable with respect
to a terminology \Term and a role hierarchy $\R$, then $C,\Term$ have
a \emph{connected} model, i. e., a model where any two elements are
connect by a role path over those roles occuring in $C$ and $\Term$, and (b)
if $y$ is reachable from $x$ via a role path (possibly involving
inverse roles), then $\tuple{x}{y} \in U\ifunc$.  These are easy
consequences of the semantics and the definition of $U$.

\begin{theorem}\label{theorem:internal}
  Satisfiability and subsumption of \shiq-concepts w.r.t. terminologies and role
  hierarchies are polynomially reducible to (un)sat\-is\-fi\-abil\-i\-ty of
  \shiq-concepts w.r.t.  role hierarchies, and therefore to consistency of
  \shiq-Aboxes w.r.t.  role hierarchies.
  
  Consistency of \shiq-Aboxes w.r.t. terminologies and role hierarchies is
  polynomially reducible to consistency of \shiq-Aboxes w.r.t.  role
  hierarchies.
\end{theorem}


\section{A \shiq-Abox Tableau Algorithm}

With Theorem~\ref{theorem:internal}, all standard inference problems for
\shiq-concepts and Aboxes can be reduced to Abox-consistency w.r.t.  a role
hierarchy.  In the following, we present a tableau-based algorithm that decides
consistency of \shiq-Aboxes w.r.t. role hierarchies, and therefore all other
\shiq inference problems presented.

The algorithm tries to construct, for a \shiq-Abox \A, a tableau for \A, that
is, an abstraction of a model of \A. Given the notion of a tableau, it is then
quite straightforward to prove that the algorithm is a decision procedure for
Abox consistency.

\subsection{A Tableau for Aboxes}

In the following, if not stated otherwise, $C,D$ denote \shiq-concepts, $\R$ a
role hierarchy, \A an Abox, $\Roles_\A$ the set of roles occurring in $\A$ and
$\R$ together with their inverses, and $\Inds_\A$ is the set of individuals
occurring in $\A$.

Without loss of generality, we assume all concepts $C$ occurring in assertions
$a\mycolon C\in \A$ to be in NNF, that is, negation occurs in front of concept
names only. Any \shiq-concept can easily be transformed into an equivalent one
in NNF by pushing negations inwards using a combination of DeMorgan's laws and
the following equivalences:
$$\begin{array}{rclrcl} \neg (\exists R.C)&\equiv& (\forall R. \neg C)&
  \neg (\forall R.C)&\equiv& (\exists R. \neg C)\\
  \neg (\atmostq n R C) &\equiv& \atleastq {(n+1)} R C&
  \neg (\atleastq n R C) &\equiv& \atmostq {(n-1)} R C \quad  \text{where}\\
  & & & \atmostq {(-1)} R C &:=& A \sqcap \neg A \quad \text{for some $A \in
  \Concepts$}
\end{array}$$
For a concept $C$ we will denote the NNF of $\neg C$ by $\nneg C$.  Next, for a
concept $C$, $\clos(C)$ is the smallest set that contains $C$ and is closed
under sub-concepts and $\nneg$. We use $\clos(\A):= \bigcup_{a\mycolon C\in \A}
\clos(C)$ for the closure $\clos(C)$ of each concept $C$ occurring in \A.  It is
not hard to show that the size of $\clos(\A)$ is polynomial in the size of \A.

\begin{definition} \label{def:tableau}
  $T = (\mathbf{S},\Lab,\Edges,\Indmap)$ is a \emph{tableau} for $\A$ w.r.t. $\R$
  iff
  \begin{itemize}
  \item $\mathbf{S}$ is a non-empty set,
  \item $\Lab:\mathbf{S} \rightarrow 2^{\clos(\A)}$ maps each element in
    $\mathbf{S}$ to a set of concepts,
  \item $\Edges:\Roles_\A \rightarrow 2^{\mathbf{S} \times \mathbf{S}}$ maps
    each role to a set of pairs of elements in $\mathbf{S}$, and
  \item $\Indmap:\Inds_\A\rightarrow \mathbf{S}$ maps individuals occurring in
    $\A$ to elements in $\mathbf{S}$.
  \end{itemize}
  Furthermore, for all $s,t \in \mathbf{S}$, $C, C_1, C_2\in \clos(\A)$, and
  $R,S \in \Roles_\A$, $T$ satisfies:
  \begin{enumerate}\setlength{\itemsep}{0ex}
  \item[\tabl 1] if $C \in \Lab(s)$, then $\neg C \notin \Lab(s)$,
  \item[\tabl 2] if $C_1 \sqcap C_2 \in \Lab(s)$, then $C_1 \in \Lab(s)$ and $C_2 \in
    \Lab(s)$,
  \item[\tabl 3] if $C_1 \sqcup C_2 \in \Lab(s)$, then $C_1 \in \Lab(s)$ or $C_2 \in
    \Lab(s)$,
  \item[\tabl 4] if $\all{S}{C} \in \Lab(s)$ and $\tuple{s}{t} \in \Edges(S)$, then $C \in
    \Lab(t)$,
  \item[\tabl 5] if $\some{S}{C} \in \Lab(s)$, then there is some $t \in \mathbf{S}$ such
    that $\tuple{s}{t} \in \Edges(S)$ and $C \in \Lab(t)$,
  \item[\tabl 6] if $\all{S}{C} \in \Lab(s)$ and $\tuple{s}{t} \in \Edges(R)$ for some $R
    \sss S$ with $\Tr(R)$, then $\all{R}{C} \in \Lab(t)$,
  \item[\tabl 7] $\tuple{x}{y}\in \Edges(R)$ iff $\tuple{y}{x}\in \Edges(\Inv(R))$,
  \item[\tabl 8] if $\tuple s t \in \Edges(R)$ and $R \sss S$, then $\tuple s t \in
    \Edges(S)$,
  \item[\tabl 9] if $\atmostq n S C \in \Lab(s)$, then $\sharp S^T(s,C) \leqslant n$,
  \item[\tabl{10}] if $\atleastq n S C \in \Lab(s)$, then $\sharp S^T(s,C)
    \geqslant n$,
  \item[\tabl{11}] if $(\bowtie \; n \; S \; C) \in \Lab(s)$ and $\tuple s t \in
    \Edges(S)$ then $C \in \Lab(t)$ or $\nneg C \in \Lab(t)$,
  \item[\tabl{12}] if $a\mycolon C\in \A$, then $C\in \Lab(\Indmap(a))$,
  \item[\tabl{13}] if $(a,b)\mycolon R\in \A$, then $\tuple{\Indmap(a)}{\Indmap(b)}\in
    \Edges(R)$,
  \item[\tabl{14}] if $a \ndoteq b \in \A$, then $\Indmap(a) \neq \Indmap(b)$,
  \end{enumerate}
  where $\bowtie$ is a place-holder for both $\leqslant$ and $\geqslant$, and $S^T(s,C) := \{ t \in \mathbf{S} \mid \tuple s t \in \Edges(S) \ 
  \text{and} \ C \in \Lab(t) \}$.
\end{definition}

\begin{lemma}\label{lemma:sat-tabl}
  A \shiq-Abox $\A$ is consistent w.r.t. $\R$ iff there exists a tableau for
  $\A$ w.r.t. $\R$.
\end{lemma}

\paragraph{Proof:} For the \emph{if} direction, if $T = (\mathbf{S},\Lab,\Edges,\Indmap)$ is a
tableau for $\A$ w.r.t. $\R$, a model $\mathcal{I} = (\deltai, \cdot\ifunc)$ of
$\A$ and $\R$ can be defined as follows:
$$
\begin{array}{r@{\quad}rcl} 
  & \deltai & := & \mathbf{S} \\[0.5ex]
  \mbox{for concept names A in
  $\clos(\A):$} & A\ifunc & := & \{s \mid A \in \Lab(s)\}  \\[0.5ex]
  \mbox{ for individual names } a \in \Inds: &  a\ifunc & := & \Indmap(a)   \\[0.5ex]
  \mbox{ for role names } R \in \R: &  R\ifunc & := &  %
    \begin{cases}
      \Edges(R)^+ & \mbox{if $\Tr(R)$} \\
      \Edges(R)\cup \bigcup\limits_{P\sss R, P\neq R}P\ifunc & \mbox{otherwise}
    \end{cases}
\end{array}$$
where $\Edges(R)^+$ denotes the transitive closure of $\Edges(R)$.  The
interpretation of non-transitive roles is recursive in order to correctly
interpret those non-transitive roles that have a transitive sub-role. From the
definition of $R\ifunc$ and \tabl 8, it follows that, if $\tuple{s}{t}\in
S\ifunc$, then either $\tuple{s}{t}\in \Edges(S)$ or there exists a path
$\tuple{s}{s_1}, \tuple{s_1}{s_2},\ldots,$ $\tuple{s_n}{t}\in \Edges(R)$ for
some $R$ with $\Tr(R)$ and $R\sss S$.

Due to \tabl 8 and by definition of $\I$, we have that $\I$ is a model of \R. 

To prove that $\I$ is a model of \A, we show that $C \in \Lab(s)$ implies $ s
\in C\ifunc$ for any $s \in \mathbf{S}$. Together with \tabl{12}, \tabl{13}, and
the interpretation of individuals and roles, this implies that $\I$ satisfies
each assertion in $\A$.  This proof can be given by induction on the length $\|
C \|$ of a concept $C$ in NNF, where we count neither negation nor integers in
number restrictions.
The only interesting case is $C = \all S E$: let $t \in \mathbf{S}$ with
$\tuple s t \in S^\I$.  There are two possibilities:
\begin{itemize}
\item $\tuple s t \in \Edges(S)$. Then \tabl 4 implies $E \in \Lab(t)$.
\item $\tuple s t \not \in \Edges(S)$. Then there exists a path
  $\tuple{s}{s_1}, \tuple{s_1}{s_2},\ldots,$ $\tuple{s_n}{t}\in \Edges(R)$ for
  some $R$ with $\Tr(R)$ and $R\sss S$. Then \tabl 6 implies $\all R E \in
  \Lab(s_i)$ for all $1 \leq i \leq n$, and \tabl 4 implies $E \in \Lab(t)$.
\end{itemize}
In both cases, $t \in E^\I$ by induction and hence $s \in C^\I$.

\bigskip

For the converse, for $\mathcal{I} = (\deltai, \cdot\ifunc)$ a model of $\A$
w.r.t. $\R$, we define a tableau $T = (\mathbf{S},\Lab,\Edges,\Indmap)$ for $\A$
and \R as follows:
$$
\mathbf{S} := \deltai, \quad \Edges(R) := R\ifunc, \quad \Lab(s) := \{C \in
\clos(\A) \mid s \in C\ifunc \}, \quad \mbox{ and } \quad \Indmap(a) =  a\ifunc.$$
It is easy to demonstrate that $T$ is a tableau for $D$. \qed

\subsection{The Tableau  Algorithm}

In this section, we present a completion algorithm that tries to construct, for
an input Abox \A and a role hierarchy \R, a tableau for \A w.r.t. \R. We prove
that this algorithm constructs a tableau for \A and \R iff there exists a
tableau for \A and \R, and thus decides consistency of \shiq Aboxes w.r.t. role hierarchies.

Since Aboxes might involve several individuals with arbitrary role relationships
between them, the completion algorithm works on a \emph{forest} rather than on a
\emph{tree}, which is the basic data structure for those completion algorithms
deciding satisfiability of a concept. Such a forest is a collection of trees
whose root nodes correspond to the individuals present in the input Abox. In the
presence of transitive roles, \emph{blocking} is employed to ensure termination
of the algorithm. In the additional presence of inverse roles, blocking is
\emph{dynamic}, i.e., blocked nodes (and their sub-branches) can be un-blocked
and blocked again later.  In the additional presence of number restrictions,
\emph{pairs} of nodes are blocked rather than single nodes. 

\begin{definition}\label{def:shin-algo}
  A \emph{completion forest} $\for$ for a \shiq Abox $\A$ is a collection of
  trees whose distinguished root nodes are possibly connected by edges in an
  arbitrary way.  Moreover, each node $x$ is labelled with a set $\Lab(x)
  \subseteq \clos(\A)$ and each edge $\tuple{x}{y}$ is labelled with a set
  $\Lab(\tuple{x}{y}) \subseteq \R_\A$ of (possibly inverse) roles occurring in
  \A. Finally, completion forests come with an explicit inequality relation
  $\ndoteq$ on nodes and an explicit equality relation $\doteq$ which are
  implicitly assumed to be symmetric.
  
  If nodes $x$ and $y$ are connected by an edge $\tuple{x}{y}$ with $R \in
  \Lab(\tuple x y)$ and $R\sss S$, then $y$ is called an $S$-\emph{successor} of
  $x$ and $x$ is called an $\Inv(S)$-\emph{predecessor} of $y$. If $y$ is an
  $S$-successor or an $\Inv(S)$-predecessor of $x$, then $y$ is called an
  $S$-neighbour of $x$. A node $y$ is a successor (resp. predecessor or
  neighbour) of $y$ if it is an $S$-successor (resp. $S$-predecessor or
  $S$-neighbour) of $y$ for some role $S$. Finally, \emph{ancestor} is the
  transitive closure of \emph{predecessor}.

  For a role $S$, a concept $C$ and a node $x$ in \for we define $S^\for(x,C)$
  by
  \[
  S^\for(x,C) := \{ y \mid \text{$y$ is $S$-neighbour of $x$ and $C \in
  \Lab(y)$} \}.
  \]
  
  A node is \emph{blocked} iff it is not a root node and it is either directly
  or indirectly blocked.  A node $x$ is \emph{directly blocked} iff none of its
  ancestors are blocked, and it has ancestors $x'$, $y$ and $y'$ such that
  \begin{enumerate}\setlength{\itemsep}{0ex}
  \item $y$ is not a root node \emph{and}
  \item $x$ is a successor of $x'$ and $y$ is a successor of $y'$ \emph{and}
  \item $\Lab(x) = \Lab(y)$ and $\Lab(x') = \Lab(y')$ \emph{and}
  \item $\Lab(\tuple{x'}{x}) = \Lab(\tuple{y'}{y})$.
  \end{enumerate}
  In this case we will say that $y$ \emph{blocks} $x$. 
  
  A node $y$ is \emph{indirectly blocked} iff one of its ancestors is blocked,
  or it is a successor of a node $x$ and $\Lab(\tuple{x}{y}) = \emptyset$; the
  latter condition avoids wasted expansions after an application of the
  $\leqslant$-rule.
  
  Given a \shiq-Abox $\A$ and a role hierarchy $\R$, the algorithm initialises a
  completion forest $\for_\A$ consisting only of root nodes. More precisely,
  $\for_\A$ contains a root node $x_0^i$ for each individual $a_i\in \Inds_\A$
  occurring in $\A$, and an edge $\tuple{x_0^i}{x_0^j}$ if \A contains an
  assertion $(a_i,a_j)\mycolon R$ for some $R$. The labels of these nodes and
  edges and the relations $\ndoteq$ and $\doteq$ are initialised as follows:
  \[
  \begin{array}{rcl}
    \Lab(x_0^i) & := & \{C \mid a_i\mycolon C \in \A\},\\
    \Lab(\tuple{x_0^i}{x_0^j}) & := & \{R \mid (a_i,a_j):R\in \A\},\\
    x_0^i \ndoteq x_0^j &\text{ iff }& a_i \ndoteq a_j \in \A\mbox{, and}
  \end{array}
  \]
  the $\doteq$-relation is initialised to be empty. $\for_\A$ is then expanded
  by repeatedly applying the rules from Figure~\ref{abox:shiq-rules}.
  
  For a node $x$, $\Lab(x)$ is said to contain a \emph{clash} if, for some
  concept name $A \in \Concepts$, $\{A, \neg A\} \subseteq \Lab(x)$, or if there
  is some concept $\atmostq n S C \in \Lab(x)$ and $x$ has $n+1$ $S$-neighbours
  $y_0,\dots,y_n$ with $C \in \Lab(y_i)$ and $y_i \ndoteq y_j$ for all $0 \leq i
  < j \leq n$. A completion forest is \emph{clash-free} if none of its nodes
  contains a clash, and it is \emph{complete} if no rule from
  Figure~\ref{abox:shiq-rules} can be applied to it.
  
  For a \shiq-Abox $\A$, the algorithm starts with the completion forest
  $\for_\A$. It applies the expansion rules in Figure~\ref{abox:shiq-rules},
  stopping when a clash occurs, and answers ``$\A$ is consistent w.r.t. $\R$''
  iff the completion rules can be applied in such a way that they yield a
  complete and clash-free completion forest, and ``$\A$ and is inconsistent
  w.r.t. $\R$'' otherwise.
\end{definition}

\begin{figure}
  \begin{center}
      \begin{tabular}{lll}
        
        $\sqcap$-rule: & if \hfill 1. & $C_1 \sqcap C_2 \in \Lab(x)$, $x$ is not
        indirectly blocked, and \\
        & \hfill 2. & $\{C_1,C_2\} \not\subseteq \Lab(x)$ \\
        
        & then & $\Lab(x) \longrightarrow \Lab(x) \cup \{C_1,C_2\}$ \\ \hline 
        
        $\sqcup$-rule: & if \hfill 1. & $C_1 \sqcup C_2 \in \Lab(x)$, $x$ is not
        indirectly blocked, and \\
        & \hfill 2. & $\{C_1,C_2\} \cap \Lab(x) = \emptyset$ \\
        
        & then & $\Lab(x) \longrightarrow \Lab(x) \cup \{E\}$
        for some $E\in \{C_1,C_2 \}$\\ \hline 
        
        $\exists$-rule: & if \hfill 1. & $\some{S}{C} \in \Lab(x)$, $x$ is not
        blocked, and \\
        & \hfill 2. & $x$ has no $S$-neighbour $y$ with $C\in \Lab(y)$\\
        
        & then & create a new node  $y$  with $\Lab(\tuple{x}{y}):=\{S\}$
        and $\Lab(y):=\{C\}$ \\\hline  
        
        $\forall$-rule: & if \hfill 1. & $\all{S}{C} \in \Lab(x)$, $x$ is not
        indirectly blocked, and \\
        & \hfill 2. &there is an $S$-neighbour $y$ of $x$ with $C \notin \Lab(y)$ \\
        
        & then & $\Lab(y) \longrightarrow \Lab(y) \cup \{C\}$ \\ \hline 
        
        $\forall_+$-rule: & if \hfill 1. & $\all{S}{C} \in \Lab(x)$,  $x$ is not
        indirectly blocked, and\\  
        & \hfill 2. & there is some $R$ with  $\Tr(R)$ and $R\sss S$, \\
        & \hfill 3. &there is an  $R$-neighbour $y$ of $x$ with $\all{R}{C}
        \notin \Lab(y)$ \\ 
        & then & $\Lab(y) \longrightarrow \Lab(y) \cup \{\all{R}{C}\}$\\ \hline 

        \textit{choose}-rule: & if \hfill 1. & $(\bowtie \; n \; S \; C) \in \Lab(x)$,
        $x$ is not indirectly blocked, and\\
        & \hfill 2. & there is an $S$-neighbour $y$ of $x$ with $\{ C,\nneg C\}
        \cap  \Lab(y)= \emptyset$\\
        
        & then & $\Lab(y) \longrightarrow \Lab(y) \cup \{E\}$
        for some $E \in \{C,\nneg C \}$\\ \hline

        $\geqslant$-rule: & if \hfill 1. & $\atleastq n S C \in \Lab(x)$, $x$ is
        not
        blocked, and\\
        & \hfill 2. & there are no $n$ $S$-neighbours $y_1,\dots,y_n$ such that
        $C \in
        \Lab(y_i)$\\
        & & and $y_i \ndoteq y_j$ for $1 \leq i < j \leq n$\\
        
        & then & create $n$ new nodes $y_1,\dots,y_n$ with $\Lab(\tuple x {y_i}) =
        \{S\}$,\\ 
        & & $\Lab(y_i) = \{C\}$, and $y_i \ndoteq y_j$ for $1 \leq i < j \leq
        n$.\\ \hline 
        
        $\leqslant$-rule: & if \hfill 1. & $\atmostq n S C \in \Lab(x)$, $x$ is
        not indirectly
        blocked, and\\
        & \hfill 2. & $\sharp S^\for(x,C) > n$, there are $S$-neighbours
        $y,z$ of $x$ with not $y \ndoteq z$, \\
        & & $y$ is neither a root node nor an ancestor of $z$, and $C \in \Lab(y)
        \cap \Lab(z)$,   \\  

        &then & 1.\ $\Lab(z) \longrightarrow \Lab(z) \cup \Lab(y)$ and\\
        & & 2.\ \begin{tabular}[t]{rl}
          \multicolumn{2}{l}{if $z$ is an ancestor of $x$} \\
          then & 
          $\begin{array}[t]{rcl}
            \Lab (\tuple{z}{x})
            &\longrightarrow&  \Lab (\tuple{z}{x}) \cup \Inv(\Lab (\tuple{x}{y}))
          \end{array}$\\
          else \hfill &
          $\begin{array}[t]{rcl}
            \Lab (\tuple{x}{z})
            &\longrightarrow&  \Lab (\tuple{x}{z}) \cup \Lab (\tuple{x}{y})
          \end{array}$
        \end{tabular}\\
        & & 3.\ $\Lab(\tuple{x}{y}) \longrightarrow \emptyset$ \\
        & & 4.\ Set $u \ndoteq z$ for all $u$ with $u \ndoteq y$\\\hline

        $\leqslant_r$-rule: & if \hfill 1. & $\atmostq n S C \in \Lab(x)$,
        and\\
        & \hfill 2. & $\sharp S^\for(x,C) > n$ and there are two $S$-neighbours
        $y,z$ of $x$ \\ & & which are both root nodes,
        $C \in \Lab(y) \cap \Lab(z)$, and not $y \ndoteq z$\\
        &then & 1.\ $\Lab(z) \longrightarrow \Lab(z) \cup \Lab(y)$ and\\
        & & 2.\ For all edges $\tuple y w$:\\
        & & \hspace{2ex} i. if the edge $\tuple z w$ does not
        exist, create it with $\Lab(\tuple z w) := \emptyset$\\
        & & \hspace{2ex} ii. $\Lab (\tuple{z}{w})
        \longrightarrow \Lab (\tuple{z}{w}) \cup \Lab (\tuple{y}{w})$ \\
        & & 3.\ For all edges $\tuple w y$:\\
        & & \hspace{2ex} i. if the edge $\tuple w z$ does not
        exist, create it with $\Lab(\tuple w z) := \emptyset$\\
        & & \hspace{2ex} ii. $\Lab (\tuple{w}{z})
        \longrightarrow \Lab (\tuple{w}{z}) \cup \Lab (\tuple{w}{y})$ \\
        & & 4.\ Set $\Lab(y):= \emptyset $ and remove all edges to/from $y$.\\
        & & 5.\ Set $u \ndoteq z$ for all $u$ with $u \ndoteq y$ and set $y
        \doteq z$.
        \vspace{-1ex}
      \end{tabular}
  \end{center} 
  \caption{The Expansion Rules for $\shiq$-Aboxes.}
  \label{abox:shiq-rules}
\end{figure}

Since both the $\leqslant$-rule and the $\leqslant_r$-rule are rather
complicated, they deserve some more explanation. Both rules deal with the
situation where a concept $\atmostq n R C \in \Lab(x)$ requires the
identification of two $R$-neighbours $y,z$ of $x$ that contain $C$ in their
labels. Of course, $y$ and $z$ may only be identified if $y \ndoteq z$ is not
asserted. If these conditions are met, then one of the two rules can be applied.
The $\leqslant$-rule deals with the case where at least one of the nodes to be
identified, namely $y$, is not a root node, and this can lead to one of two
possible situations, both shown in Figure~\ref{fig:effect-leq-rule}.
\begin{figure}[t]
    \framebox[\textwidth]{ 
      ~\hspace{-0.8cm}\input{leq-rule.pstex_t}}
  \caption{Effect of the $\leqslant$- and the $\leqslant_r$-rule}
  \label{fig:effect-leq-rule}
\end{figure}
The upper situation occurs when both $y$ and $z$ are successors of $x$. In this
case, we add the label of $y$ to that of $z$, and the label of the edge $\tuple
x y$ to the label of the edge $\tuple x z$.  Finally, $z$ inherits all
inequalities from $y$, and $\Lab(\tuple x y)$ is set to $\emptyset$, thus
blocking $y$ and all its successors.

The second situation occurs when both $y$ and $z$ are neighbours of $x$, but $z$
is the predecessor of $x$.  Again, $\Lab(y)$ is added to $\Lab(z)$, but in this
case the inverse of $\Lab(\tuple x y)$ is added to $\Lab(\tuple z x)$, because
the edge $\tuple x y$ was pointing away from $x$ while $\tuple z x$ points
towards it.  Again, $z$ inherits the inequalities from $y$ and $\Lab(\tuple x
y)$ is set to $\emptyset$.

The $\leqslant_r$ rule handles the identification of two root nodes. An example
of the whole procedure is given in the lower part of Figure~\ref{fig:effect-leq-rule}.
In this case, special care has to be taken to preserve the relations introduced
into the completion forest due to role assertions in the Abox, and to memorise
the identification of root nodes (this will be needed in order to construct a
tableau from a complete and clash-free completion forest). The $\leqslant_r$
rule includes some additional steps that deal with these issues.  Firstly, as
well as adding $\Lab(y)$ to $\Lab(z)$, the edges (and their respective labels)
between $y$ and its neighbours are also added to $z$.  Secondly, $\Lab(y)$ and
all edges going from/to $y$ are removed from the forest. This will not lead to
dangling trees, because all neighbours of $y$ became neighbours of $z$ in the
previous step.  Finally, the identification of $y$ and $z$ is recorded in the
$\doteq$ relation.

\begin{lemma} \label{lemma:shin-algo-term}
  Let $\A$ be a \shiq-Abox and $\R$ a role hierarchy.  The completion algorithm
  terminates when started for $\A$ and $\R$.
\end{lemma}

\paragraph{Proof:} Let $m = \sharp\clos(\A) $, $n =
\Card{\Roles_\A}$, and $n_{\max}:=\max\{ n\mid \atleastq n R C \in \clos(\A)\}$.
Termination is a consequence of the following properties of the expansion rules:
\begin{enumerate}
\item The expansion rules never remove nodes from the forest. The only rules that
  remove elements from the labels of edges or nodes are the $\leqslant$- and
  $\leqslant_r$-rule, which sets them to $\emptyset$. If an edge label is set to
  $\emptyset$ by the $\leqslant$-rule, the node below this edge is blocked and
  will remain blocked forever. The $\leqslant_r$-rule only sets the label of a
  root node $x$ to $\emptyset$, and after this, $x$'s label is never changed
  again since all edges to/from $x$ are removed.  Since no root nodes are
  generated, this removal may only happen a finite number of times, and the new
  edges generated by the $\leqslant_r$-rule guarantees that the resulting
  structure is still a completion forest.
\item Nodes are labelled with subsets of $\clos(\A)$ and edges with subsets of
  $R_\A$, so there are at most $2^{2mn}$ different possible labellings for a
  pair of nodes and an edge.  Therefore, if a path $p$ is of length at least
  $2^{2mn}$, the pair-wise blocking condition implies the existence of two nodes
  $x,y$ on $p$ such that $y$ directly blocks $y$. Since a path on which nodes
  are blocked cannot become longer, paths are of length at most $2^{2mn}$. 
\item Only the $\exists$- or the $\geqslant$-rule generate new nodes, and each
  generation is triggered by a concept of the form \some{R}{C} or $\atleastq n R
  C$ in $\clos(\A)$. Each of these concepts triggers the generation of at most
  $n_{\max}$ successors $y_i$: note that if the $\leqslant$- or the
  $\leqslant_r$-rule subsequently causes $\Lab(\tuple{x}{y_i})$ to be changed to
  $\emptyset$, then $x$ will have some $R$-neighbour $z$ with $\Lab(z) \supseteq
  \Lab(y)$. This, together with the definition of a clash, implies that the rule
  application which led to the generation of $y_i$ will not be repeated. Since
  $\clos(\A)$ contains a total of at most $m$ \some{R}{C}, the out-degree of the
  forest is bounded by $m n_{\max} n $.  \qed
\end{enumerate}

\begin{lemma} \label{lemma:shin-algo-correct}
  Let $\A$ be a \shiq-Abox and $\R$ a role hierarchy. If the expansion rules can
  be applied to $\A$ and $\R$ such that they yield a complete and clash-free
  completion forest, then $\A$ has a tableau w.r.t. $\R$.
\end{lemma}

\paragraph{Proof:} 
Let \for be a complete and clash-free completion forest. The definition of a tableau $T
= (\mathbf{S},\Lab,\Edges, \Indmap)$ from \for works as follows. Intuitively, an
individual in $\mathbf{S}$ corresponds to a {\em path} in \for from some root
node to some node that is not blocked, and which goes only via non-root nodes.
  
More precisely, a \emph{path} is a sequence of pairs of nodes of \for of the
form $p = [\pair{x_0}{x'_0}, \dots, \pair{x_n}{x'_n}]$. For such a path we
define $\Tail(p) := x_n$ and $\Tail'(p) := x'_n$. With
$[p|\pair{x_{n+1}}{x'_{n+1}}]$, we denote the path $[\pair{x_0}{x'_0}, \dots,
\pair{x_n}{x'_n}, \pair{x_{n+1}}{x'_{n+1}}]$. The set $\Paths(\for)$ is defined
inductively as follows:
\begin{itemize}
\item For root nodes $x_0^i$ of \for, $[\pair{x_0^i}{x_0^i}] \in \Paths(\for)$,
  and
\item For a path $p \in \Paths(\for)$ and a node $z$ in \for:
  \begin{itemize}
  \item if $z$ is a successor of $\Tail(p)$ and $z$ is neither blocked nor a
    root node, then $[p|\pair{z}{z}] \in \Paths(\for)$, or
  \item if, for some node $y$ in \for, $y$ is a successor of $\Tail(p)$ and $z$
    blocks $y$, then $[p|\pair{z}{y}] \in \Paths(\for)$.
  \end{itemize}
\end{itemize}
Please note that, since root nodes are never blocked, nor are they blocking
other nodes, the only place where they occur in a path is in the first place.
Moreover, if $p \in \Paths(\for)$, then $\Tail(p)$ is not blocked, $\Tail(p) =
\Tail'(p)$ iff $\Tail'(p)$ is not blocked, and $\Lab(\Tail(p)) =
\Lab(\Tail'(p))$.

We define a tableau $T = (\mathbf{S},\Lab,\Edges, \Indmap)$ as follows:
\[
\begin{array}{r@{\,}c@{\,}l}
  \mathbf{S} & = & \Paths(\for)\\[0.5ex]
  \Lab(p) & = & \Lab(\Tail(p))\\[0.5ex]
  \Edges(R) & = & \{\tuple{p}{[p|\pair{x}{x'}]} \in \mathbf{S} \times
  \mathbf{S} \mid
  \text{$x'$ is an $R$-successor of $\Tail(p)$}\} \cup\mbox{}\\
    & & \{\tuple{[q|\pair{x}{x'}]}{q} \in \mathbf{S} \times \mathbf{S} \mid
    \text{$x'$ is an $\Inv(R)$-successor of $\Tail(q)$}\}  \cup\mbox{}\\
    & & \{\tuple{\path{\pair{x}{x}}}{\path{\pair{y}{y}}} \in \mathbf{S}
    \times \mathbf{S} \mid
    x, y \mbox{ are root nodes, and }
    y \mbox{ is an $R$-neighbour of  } x \} \\
    \Indmap(a_i)&=& \left \{\begin{array}[c]{ll}
        \path{\pair{x_0^i}{x_0^i}}&\text{ if } x_0^i \text{ is a root node in
        \for with
        }\Lab(x_0^i)\not = \emptyset \\
        \path{\pair{x_0^j}{x_0^j}}&\text{ if } \Lab(x_0^i)=\emptyset,
        x_0^j\text{ a root node in \for with } \Lab(x_0^j) \neq \emptyset \text{
        and } x_0^i \doteq x_0^j
  \end{array}\right .
\end{array}
\]           
\noindent Please note that $\Lab(x)=\emptyset$ implies that $x$ is a root node
and that there is another root node $y$ with $\Lab(y)\not = \emptyset$ and $x \doteq
y$. We show that $T$ is a tableau for $D$.
\begin{itemize}
\item $T$ satisfies \btabl 1 because \for is clash-free.
  
\item \btabl 2 and \btabl 3 are satisfied by $T$ because \for is complete.
  
\item For \btabl 4, let $p,q\in \mathbf{S}$ with $\all{R}{C} \in \Lab(p)$,
  $\tuple{p}{q}\in \Edges(R)$. If $q = \path{p|\pair{x}{x'}}$, then $x'$ is an
  $R$-successor of $\Tail(p)$ and, due to completeness of \for, $C\in \Lab(x')=
  \Lab(x)=\Lab(q)$.
  If $p = \path{q|\pair{x}{x'}}$, then $x'$ is an $\Inv(R)$-successor of
  $\Tail(q)$ and, due to completeness of \for, $C\in \Lab(\Tail(q))= \Lab(q)$.
    If $p =\path{\pair x x}$ and $q=\path{\pair y y}$ for two root nodes $x$, $x$,
  then $y$ is an $R$-neighbour of $x$, and completeness of \for yields $C\in
  \Lab(y) =\Lab(q)$.
  \btabl 6 and \btabl{11} hold for similar reasons.
  
\item For \btabl 5, let $\some{R}{C} \in \Lab(p)$ and $\Tail(p) = x$. Since $x$
  is not blocked and \for complete, $x$ has some $R$-neighbour $y$ with $C\in
  \Lab(y)$.
  \begin{itemize}
  \item If $y$ is a successor of $x$, then $y$ can either be a root node or
    not. 
    \begin{itemize}
    \item If $y$ is not a root node: if $y$ is not blocked, then $q :=
      [p|\pair{y}{y}] \in \mathbf{S}$; if $y$ is blocked by some node $z$,
      then $q := [p|\pair{z}{y}] \in \mathbf{S}$.
    \item If $y$ is a root node: since $y$ is a successor of $x$, $x$ is also a
      root node. This implies $p = \path{\pair x x }$ and $q = \path{\pair y y}
      \in \mathbf{S}$.
  \end{itemize}
  \item $x$ is an $\Inv(R)$-\emph{successor} of $y$, then either
    \begin{itemize}
    \item $p = [q|\pair{x}{x'}]$ with $\Tail(q) = y$.
    \item $p = [q|\pair{x}{x'}]$ with $\Tail(q) = u \neq y$.  Since $x$ only has
      one predecessor, $u$ is not the predecessor of $x$.  This implies $x \neq
      x'$, $x$ blocks $x'$, and $u$ is the predecessor of $x'$ due to the
      construction of $\Paths$.  Together with the definition of the blocking
      condition, this implies $\Lab(\tuple u {x'}) = \Lab(\tuple y x)$ as well
      as $\Lab(u) = \Lab(y)$ due to the blocking condition.
    \item $p=[\pair x x]$ with $x$ being a root node. Hence
      $y$ is also a root node and $q = [\pair y y]$.
    \end{itemize}
  \end{itemize}
  In any of these cases, $\tuple p q \in \Edges(R)$ and $C \in \Lab(q)$.

\item \btabl 7 holds because of the symmetric definition of the mapping $\Edges$.
  
\item \btabl 8 is due to the definition of $R$-neighbours and $R$-successor.
    
\item Suppose \btabl 9 were not satisfied. Hence there is some $p \in \mathbf{S}$
  with $(\atmostq n S C) \in \Lab(p)$ and $\sharp S^T(p,C)  > n$. We will show that
  this implies $\sharp S^\for(\Tail(p),C) > n$, contradicting either
  clash-freeness or completeness of \for.
  Let $x := \Tail(p)$ and $P := S^T(p,C)$. We distinguish two cases:
  \begin{itemize}
  \item $P$ contains only paths of the form $[p|\pair{y}{y'}]$ and
    $\path{\pair{x_0^{i_\ell}}{x_0^{i_\ell}}}$. Then $\sharp P > n$ is
    impossible since the function $\Tail'$ is injective on $P$: if we assume
    that there are two distinct paths $q_1,q_2 \in P$ and $\Tail'(q_1) =
    \Tail'(q_2) = y'$, then this implies that each $q_i$ is of the form $q_i =
    [p | \pair{y_i}{y'}]$ or $q_i = [\pair{y'}{y'}]$.  From $q_1 \neq q_2$, we
    have that $q_i = [p | \pair{y_i}{y'}]$ holds for some $i\in \{1,2\}$. Since root
    nodes occur only in the beginning of paths and $q_1\not = q_2$, we have $q_1
    = [p | (y_1,y')]$ and $q_2 = [p | (y_2,y')]$.  If $y'$ is not blocked, then
    $y_1 = y' = y_2$, contradicting $q_1 \neq q_2$.  If $y'$ is blocked in \for,
    then both $y_1$ and $y_2$ block $y'$, which implies $y_1 = y_2$, again a
    contradiction.
    Hence $\Tail'$ is injective on $P$ and thus $\sharp P = \sharp \Tail'(P)$.
    Moreover, for each $y' \in \Tail'(P)$, $y'$ is an $S$-successor of $x$ and
    $C \in \Lab(y')$. This implies $\sharp S^\for(x,C) > n$.
  \item $P$ contains a path $q$ where $p = [q | \pair{x}{x'}]$.  Obviously, $P$
    may only contain one such path. As in the previous case, $\Tail'$ is an
    injective function on the set $P' := P \setminus \{ q \}$, each $y' \in
    \Tail'(P')$ is an $S$-successor of $x$, and $C \in \Lab(y')$ for each $y'
    \in \Tail'(P')$. Let $z:= \Tail(q)$.
    We distinguish two cases:
    \begin{itemize}
    \item $x=x'$. Hence $x$ is not blocked, and thus $x$ is an
      $\Inv(S)$-successor of $z$. Since $\Tail'(P')$ contains only successors of
      $x$ we have that $z \not\in \Tail'(P')$ and, by construction, $z$ is an
      $S$-neighbour of $x$ with $C \in \Lab(z)$.
    \item $x \neq x'$. This implies that $x'$ is blocked by $x$ and that $x'$ is
      an $\Inv(S)$-successor of $z$. Due to the definition of pairwise-blocking
      this implies that $x$ is an $\Inv(S)$-successor of some node $u$ with
      $\Lab(u) = \Lab(z)$. Again, $u \not\in \Tail'(P')$ and, by construction,
      $u$ is an $S$-neighbour of $x$ and $C \in \Lab(u)$.
    \end{itemize}
  \end{itemize}
    
\item For \btabl{10}, let $(\atleastq n S C) \in \Lab(p)$. Hence there are $n$
  $S$-neighbours $y_1, \dots, y_n$ of $x = \Tail(p)$ in \for with $C \in
  \Lab(y_i)$.  For each $y_i$ there are three possibilities:
  \begin{itemize}
  \item $y_i$ is an $S$-successor of $x$ and $y_i$ is not blocked in \for. Then
    $q_i := [p|\pair{y_i}{y_i}]$ or $y_i$ is a root node and $q_i
    :=\path{\pair{y_i}{y_i}}$ is in $\mathbf S$.
  \item $y_i$ is an $S$-successor of $x$ and $y_i$ is blocked in \for by some
    node $z$. Then $q_i = [p|\pair{z}{y_i}]$ is in $\mathbf S$. Since the same
    $z$ may block several of the $y_j$s, it is indeed necessary to include $y_i$
    explicitly into the path to make them distinct.
  \item $x$ is an $\Inv(S)$-successor of $y_i$. There may be at most one such
    $y_i$ if $x$ is not a root node. Hence either $p = [q_i |
    \pair{x}{x'}]$ with $\Tail(q_i) = y_i$, or $p = [\pair x x]$ and $q_i =
    [\pair{y_i}{y_i}]$.
  \end{itemize}
  Hence for each $y_i$ there is a different path $q_i$ in $\mathbf{S}$ with
  $S\in \Lab(\tuple{p}{q_i})$ and $C\in \Lab(q_i)$, and thus $\sharp S^T(p,C)
  \geqslant n$.
  

\item \btabl{12} is due to the fact that, when the completion algorithm is started
  for an Abox \A, the initial completion forest $\for_\A$ contains, for each
  individual name $a_i$ occurring in $\A$, a root node $x_0^i$ with 
  $ \Lab(x_0^i) =\{C\in \clos(\A)\mid a_i\mycolon C\in \A \}.$
  The algorithm never blocks root individuals, and, for each root node $x_0^i$
  whose label and edges are removed by the $\leqslant_r$-rule, there is another
  root node $x_0^j$ with $x_0^i\doteq x_0^j$ and $\{C\in \clos(\A)\mid
  a_i\mycolon C\in \A \}\subseteq \Lab(x_0^j)$. Together with the definition of
  $\Indmap$, this yields \tabl{12}. \btabl{13} is satisfied for similar reasons.
  
\item \btabl{14} is satisfied because the $\leqslant_r$-rule does not identify two
  root nodes $x_0^i, y_0^i$ when $x_0^i \ndoteq y_0^i$ holds. \qed
\end{itemize}

\begin{lemma} \label{lemma:shin-algo-complete}
  Let $\A$ be a \shiq-Abox and $\R$ a role hierarchy.  If $\A$ has a tableau
  w.r.t. $\R$, then the expansion rules can be applied to $\A$ and $\R$ such
  that they yield a complete and clash-free completion forest.
\end{lemma}

\paragraph{Proof:}
Let $T = (\mathbf{S},\Lab,\Edges,\Indmap)$ be a tableau for $\A$ and $\R$. We
use $T$ to trigger the application of the expansion rules such that they yield a
completion forest $\for$ that is both complete and clash-free. To this purpose,
a function $\pi$ is used which maps the nodes of $\for$ to elements of
$\mathbf{S}$.
The mapping $\pi$ is defined as follows:
\begin{itemize}
\item For individuals $a_i$ in $\A$, we define $\pi(x_0^i) := \Indmap(a_i)$.
\item If $\pi(x)= s$ is already defined, and a successor $y$ of $x$ was
  generated for $\some{R}{C}\in \Lab(x)$, then $\pi(y)= t$ for some $t\in
  \mathbf{S}$ with $C\in \Lab(t)$ and $\tuple{s}{t}\in \Edges (R)$.
\item If $\pi(x)= s$ is already defined, and successors $y_i$ of $x$ were
  generated for $\atleastq n R C \in \Lab(x)$, then $\pi(y_i)= t_i$ for $n$
  distinct $t_i\in \mathbf{S}$ with $C\in \Lab(t_i)$ and $\tuple{s}{t_i}\in
  \Edges (R)$.
\end{itemize}

\noindent Obviously, the mapping for the initial completion forest for \A and \R
  satisfies the following conditions:
\begin{equation}
  \left.
    \begin{array}{l}
      \Lab(x) \subseteq \Lab(\pi(x)),\\
      \text{if $y$ is an $S$-neighbour of $x$, then $\tuple {\pi(x)} {\pi(y)}
      \in
      \Edges(S)$, and }\\
      \text{$x \ndoteq y$ implies $\pi(x) \neq \pi(y)$.}
    \end{array}
  \right \} 
  \tag{$*$}
\end{equation}
It can be shown that the following claim holds:

\noindent \textsc{Claim:} Let \for be  generated by the completion algorithm
for \A and \R and let $\pi$ satisfy $(*)$. If an expansion rule is applicable to
$\for$, then this rule can be applied such that it yields a completion forest
$\for'$ and a (possibly extended) $\pi$ that satisfy $(*)$.

As a consequence of this claim, \tabl 1, and \tabl{9}, if \A and \R have a
tableau, then the expansion rules can be applied to \A and \R such that they
yield a complete and clash-free completion forest. \qed

From Theorem~\ref{theorem:internal}, Lemma~\ref{lemma:sat-tabl},
\ref{lemma:shin-algo-term} \ref{lemma:shin-algo-correct}, and
\ref{lemma:shin-algo-complete}, we thus have the following theorem:

\begin{theorem}\label{theorem:abox-alchqi-dec}
  The completion algorithm is a decision procedure for the consistency of
  \shiq-Aboxes and the satisfiability and subumption of concepts with respect to
  role hierarchies and terminologies.
\end{theorem}

\section{Conclusion}

We have presented an algorithm for deciding the satisfiability of \shiq KBs
where the Abox may be non-empty and where the uniqueness of individual names is
not assumed but can be asserted in the Abox. This algorithm is of particular
interest as it can be used to decide the problem of conjunctive query
containment w.r.t.\ a schema~\cite{HorrocksSattler+-LTCS-99-15}.

An implementation of the \shiq Tbox satisfiability algorithm is
already available in the \Fact system~\cite{Horrocks99c}, and is able
to reason efficiently with Tboxes derived from realistic ER schemas.
This suggests that the algorithm presented here could form the basis
of a practical decision procedure for the query containment problem.
Work is already underway to test this conjecture by extending the
\Fact system with an implementation of the new algorithm.

\bibliographystyle{plain}

\end{document}
